\documentclass[conference,a4paper]{IEEEtran}
\IEEEoverridecommandlockouts
\usepackage{cite}
\usepackage{soul}
\usepackage{url}
\usepackage{amsmath,amssymb,amsfonts}
\usepackage{algorithmic}
\usepackage{graphicx}
\usepackage{textcomp}
\usepackage{xcolor}
\usepackage{pgf}
\usepackage{subcaption}
\def\BibTeX{{\rm B\kern-.05em{\sc i\kern-.025em b}\kern-.08em
    T\kern-.1667em\lower.7ex\hbox{E}\kern-.125emX}}
\usepackage{hyperref}
    
\begin{document}

\title{A Robust and Explainable Data-Driven Anomaly Detection Approach For Power Electronics}

\author{
\IEEEauthorblockN{Alexander Beattie\IEEEauthorrefmark{1}, Pavol Mulinka\IEEEauthorrefmark{2}, Subham Sahoo\IEEEauthorrefmark{3}, Ioannis T.  Christou\IEEEauthorrefmark{4}, \\  Charalampos Kalalas\IEEEauthorrefmark{2}, Daniel Gutierrez-Rojas\IEEEauthorrefmark{1}, and Pedro H. J. Nardelli\IEEEauthorrefmark{1}}

\IEEEauthorblockA{\IEEEauthorrefmark{1}LUT University, Finland\\}	
\IEEEauthorblockA{\IEEEauthorrefmark{2}Centre Tecnol\`{o}gic de Telecomunicacions de Catalunya (CTTC/CERCA), Barcelona, Spain\\}
\IEEEauthorblockA{\IEEEauthorrefmark{3}Aalborg University, Denmark\\}
\IEEEauthorblockA{\IEEEauthorrefmark{4}The American College of Greece, Athens, Greece\\}
Emails: alexander.beattie@student.lut.fi, \{pmulinka, ckalalas\}@cttc.es,  \\  sssa@energy.aau.dk, ichristou@acg.edu, \{daniel.gutierrez.rojas, pedro.nardelli\}@lut.fi}



\maketitle


\begin{abstract}
Timely and accurate detection of anomalies in power electronics is becoming increasingly critical for maintaining complex production systems. Robust and explainable strategies help decrease system downtime and preempt or mitigate infrastructure cyberattacks. 
This work begins by explaining the types of uncertainty present in current datasets and machine learning algorithm outputs. Three techniques for combating these uncertainties are then introduced and analyzed. We further present two anomaly detection and classification approaches, namely the Matrix Profile algorithm and anomaly transformer, which are applied in the context of a power electronic converter dataset. 
Specifically, the Matrix Profile algorithm is shown to be well suited as a generalizable approach for detecting real-time anomalies in streaming time-series data. The STUMPY python library implementation of the iterative Matrix Profile is used for the creation of the detector.
A series of custom filters is created and added to the detector to tune its sensitivity, recall, and detection accuracy.
Our numerical results show that, with simple parameter tuning, the detector provides high accuracy and performance in a variety of fault scenarios.
%
%
\end{abstract}


\section{Introduction}

Recent advances in power electronics have increased renewable energy generation, which positively contributes to global decarbonization goals. 
Converters are among the key power electronic components that play an important role in enabling an increase in renewable energy sources and storage units.  Photovoltaic power plants, wind farms, and electric vehicles utilize these converters so improving them is critical to increasing operational reliability.
For instance, power converters are prone to different classes of faults, including issues in physical components, faulty operation in the electrical grid and cyberattacks from external agents.

Anomaly detection strategies are thus required to enhance  security and reliability, and enable a widespread penetration of converters in the grid  \cite{he2020}. 
Most of the anomaly detection methods in the literature rely on classic strategies like switching pattern and voltage observation \cite{bec2021, zhou2018} or frequency analysis of output voltages \cite{lezana2006}.
Despite the effectiveness of these approaches, they are quite application dependent, focused mainly on the modulation techniques.  

Improvements in computational techniques and machine learning models are now opening the opportunity for anomaly detection in converters, with the advantage of generalization and no interdependence on specific parameters across the models. In \cite{xin2017}, a fault detection method based on deep neural networks using a sparse auto encoder obtained promising results. In \cite{sun2018}, a methodology based on wavelet packet decomposition is used to obtain energy values of the voltage signals which are then input into a 4-layer deep belief network to perform fault diagnosis. A long short-term memory network to identify different fault types in high-speed train converters is proposed in \cite{dong2021}, with an analysis of the model sensitivity on single-sensor and multisensor signals. Some notable contributions in categorizing between faults and cyberattacks in power electronic systems has also been carried out, using conventional and advanced  physics-informed machine learning methods in \cite{Anomaly1, Anomaly2, Anomaly3}.

Scientific research based on machine learning models for fault diagnosis in power electronics brings a lot of benefits, mostly in terms of computational time and the ability to deal with system parametrization \cite{kim2020,gan2020,ye2020}. However, most approaches lack ``explainability", which means that the outcomes from the learned model are not easily interpretable by humans. Interpretability and explainability have become a recent topic in the machine learning community, answering questions about the model outcomes and guiding researchers on how to improve and track their current algorithms \cite{xai}. Explainable artificial intelligence (XAI) methods for fault diagnosis in electric systems are proposed in \cite{GutierrezRojas2022}. XAI in power electronics can further improve interpretation, finding links that correlate sensor variables and how data transformation may impact fault diagnosis.

\begin{figure}[!t] 
		\centering
		\includegraphics[width=\columnwidth]{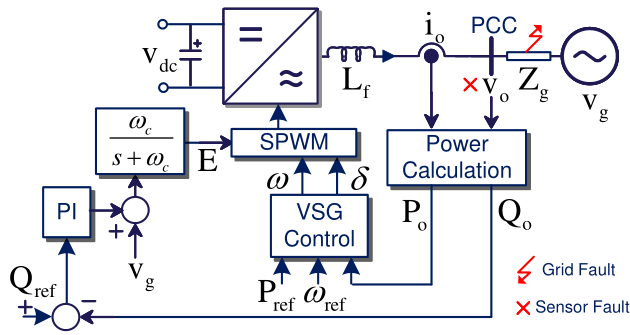}
		\caption{{Single-line diagram of a grid-forming converter interfaced to the grid via filters. We consider two classes of faults in this work - different sub-classes of faults in the grid and sensors.}}
		\label{fig:sys_model}
		\vspace{-0.4cm}
	\end{figure}

In this paper, we propose a general framework for anomaly detection, considering different types of uncertainty and anomaly in cyber-physical systems \cite{nardelli2022cyber}, which allows for testing several machine learning approaches.
To illustrate the efficacy of the proposed robust and explainable framework, we consider a simple example of the controllability of a 2-level, 3 phase grid-tied voltage source converter (VSC). As shown in Fig. \ref{fig:sys_model}, the control structure has been implemented
in the $dq0$ frame. Since this architecture involves forming the grid at its output, we use VSG control philosophy \cite{pedro} to obtain the frequency and phase angle information. As the grid-tied VSC operates with different active and reactive power reference points given by $i_{dref}$ and $i_{qref}$ respectively, the corresponding data has been obtained from the control platform to train an ensemble regression based learning model to imitate the control response. 
Our numerical results illustrate the efficacy of the proposed solution.

The remainder of the paper is divided as follows.
Section \ref{two} presents an overview of explainability in machine learning.
Section \ref{three} introduces different classes of anomaly detection approaches.
Section \ref{four} presents the numerical results for the adopted power grid model.
Section \ref{five} concludes the paper.

\section{Algorithm Explainability} \label{two}
Certain high-risk use cases for machine learning algorithms demand a high level of explainability and confidence in the algorithm outputs for decision-making. In many cases, an algorithm makes a classification decision, but there is no clear explanation as to \textit{why}. In the field of power electronics, understanding \textit{why} a data-driven controller is making a decision is critical to incorporating it into real world power distribution scenarios.

\subsection{Types of Uncertainty}
Determining sources of uncertainty in a model or system is essential. In the following, we differentiate between two basic types of uncertainty.

\textbf{Epistemic} uncertainty results from inadequate training data. In this case, the training data is not sufficient to provide enough or accurate data to the model.
This can result from imbalanced or insufficient training information.
Increasing the amount of training data or decreasing class imbalance can help reduce this type of uncertainty.

\textbf{Aleatoric} uncertainty arises from probabilistic errors in sampling that follow a specific probability distribution.
This type of uncertainty is independent of the amount of data collected and therefore cannot be corrected with additional training data.
If a signal has noise inline with a given probability distribution, having more data on that signal does not change the noise probability distribution.
This type of uncertainty references the distribution of random errors in the data and not the data distribution itself.

As an example, suppose there is a continuous audio recording at a train station.
There are three announcements each hour that disrupt the recording, but their occurrence in the hour is unknown.
Having more data (hours), does not change the amount of announcements that occur.
In the systems' domain, a malfunctioning sensor or bad connection can generate aleatoric uncertainty in the form of noise that cannot be fixed by more measurements or more data.

\subsection{Combating Model Uncertainty}
With deep learning algorithms, it is important to know the confidence level of the model output.
Authors in \cite{explaining-adversarial-examples} explain that adding simple adversarial data (e.g., similar to adding small noise to a photo) makes an image recognition algorithm incorrectly classify an animal as a completely unrelated one.
This is further concerning since adversarial data does not need to be tailored to a specific algorithm.
The transferability of this type of adversarial data allows its application against many black-box algorithms to achieve an unintended or potentially malicious result.

To solve this problem, authors in \cite{black-box-explainability} propose using conditional entropy to determine how each input is related to each output.
The generated plots are then compared to the physical insights of the system.
In the process, adversarial data that falls outside the plot is identified and removed, and the model is retrained.
This technique is helpful for identifying and removing adversarial data that falls outside the range of accepted values.
Unfortunately, it does not identify data overloading in a specific portion of the graph, which would create an incorrect classification.

\subsubsection{Bayesian Dropout}

Bayesian techniques can be used to create a probability distribution over the weights of each neuron to determine a level of prediction uncertainty.
The work in \cite{bayesian-weight-uncertainty-neural-networks} introduces a standard Bayesian neural network implementation with back-propagation that can determine these probability distributions.
%
Authors in \cite{gal2016dropout} explain that retraining a large number of models on a variety of datasets is computationally expensive and time-consuming.
A dropout technique can instead be used to approximate the Bayesian representation with improved computational efficiency.
This technique avoids over-fitting by randomly sampling and dropping network nodes across many different training iterations.
Performing Bayesian dropout while training and testing the algorithm enables the computation of variance to determine the uncertainty level of the outputs.
This allows researchers to determine if an algorithm is providing a \textit{best-guess} answer with high levels of uncertainty for specific values; in turn, this can signal the need for human intervention or review before making decisions based on the algorithm output.

\subsubsection{Shapley Additive Explanations (SHAP)}

Using reverse-engineering, the output of a machine learning algorithm can be analyzed and explained. Authors in \cite{SHAP-og-paper} introduce \textbf{SH}apley \textbf{A}dditive ex\textbf{P}lanations (SHAP) to interpret and explain the \textit{why} behind machine learning algorithm results.
Machine learning models usually output the likelihood of a certain prediction given a set of inputs. SHAP explains why a model makes a classification decision, in terms of how important each of the input features is for a given decision. To determine this, SHAP analyzes every possible combination of input weights to determine how significantly each input contributes to the overall output.

\subsubsection{Rule-based Explanations}
In the data mining literature, there exists a large body of research focusing on extracting association rules from data. Starting with the development of algorithms for the extraction of qualitative association rules \cite{apriori}, \cite{fpgrowth} of the type $bread \longrightarrow milk$, various algorithms were proposed over the years for extracting mixed and quantitative association rules that hold over a dataset \cite{brs, qarma1,qarma2}. 

Several EU-funded projects (FIREMAN, QU4LITY, AI4PublicPolicy) recently demonstrated that having all the rules that apply on a given dataset available can be very useful in explaining a third-party classifier/regressor's decision. Given a new data instance, together with the decision of the classifier, a single scan over the database of all rules that hold on the dataset can select the best-fit rule that explains the decision; a best-fit rule is one whose antecedent preconditions are satisfied by the new instance's feature values, and the rule's consequent target value matches optimally the decision made by the third-party classifier (constrains the target value as tightly as possible to an interval that contains the classifier decision value). In case of multiple rules satisfying the tightness criterion, the rule with the highest confidence and then support can break the ties; alternatively, all rules optimally supporting the decision can be presented to the user seeking explanations for the decision of a black-box classifier. 

In the context of the AI4PublicPolicy project\footnote{https://www.ai4publicpolicy.eu} research in explainable AI, a REST web-service is developed, explaining the decisions of a deep neural network predicting the number of available parking slots in the Municipality of Athens. This was performed by first extracting all rules of the form $I_1 \in [l_1, h_1] \And \dots I_k \in [l_k, h_k] \rightarrow T \geq v $ as well as all rules of the form $I_1 \in [l_1, h_1] \And \dots I_k \in [l_k, h_k] \rightarrow T \leq v $ that hold with sufficiently high minimum support and confidence thresholds on the dataset, using the QARMA algorithm \cite{qarma1}. Then, when a new HTTP POST request arrives on our web-service end point, we perform a single in-memory scan of the rulesets above and we return the best ones. 

\subsection{Anomaly Taxonomy}

\begin{figure}
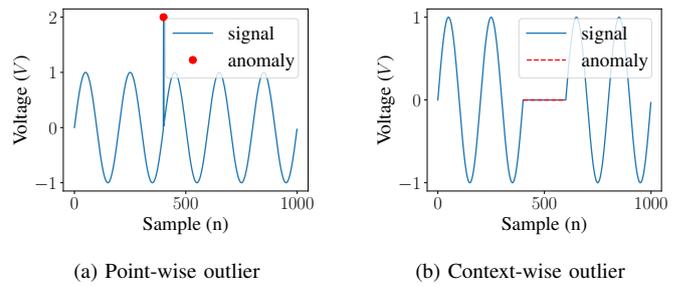

\centering
    \begin{subfigure}[b]{0.475\columnwidth}
    \centering
         {\resizebox{\textwidth}{!}{\input{"outlier_taxonomy/point_outlier.pgf"}}}
         \caption{Point-wise outlier}
         \label{fig:point}
    \end{subfigure}
    \hfill
    \begin{subfigure}[b]{0.475\columnwidth}
         \centering
          {\resizebox{\textwidth}{!}{\input{"outlier_taxonomy/contextual_outlier.pgf"}}}
         \caption{Context-wise outlier}
         \label{fig:contextual}
    \end{subfigure}
    \caption{Outlier taxonomy (adapted from \cite{lai2021revisiting}).}
    \label{fig:outliers-graphic}
    \vspace{-0.2cm}
\end{figure}

The term `outlier' or `anomaly' can have a variety of meanings depending on the context. In order to select appropriate techniques for outlier detection, it is essential to create a taxonomy for various outlier types. Fig. \ref{fig:outliers-graphic} illustrates the difference between the two types of outliers described below.

\subsubsection{Point-wise Outliers}

Point-wise outliers are single points that are anomalous in respect to the global dataset.
This can cause many problems in machine learning algorithms, and a large body of outlier detection research is focused around this area.
These outliers can be non-temporal or temporal in nature and often represent phenomena, such as intermittent sensor failure \cite{9217234}.
Additionally, such outliers may skew scaling and normalization operations.
It is important to consider the presence of point-wise outliers when selecting which type of scaler to use.
Minimum-maximum scaling is a popular choice for scaling a dataset, but it is not robust to outliers.

\subsubsection{Context-wise Outliers}

They represent a series of points that are anomalous and can be described based on their reference frame as \textit{local} and \textit{global}.
Local contextual outliers are anomalous in respect to a specific sub-set or window of the dataset.
Global contextual outliers are similar to local contextual outliers, however the window size is the global dataset.
Singular points within the global contextual outlier sub-set are usually not anomalous, but the entire pattern is.

In general, these two outlier types are treated as pattern-wise contextual outliers with differing window sizes.
These outliers often occur in time-series data because of the interdependence of samples and sampling time.
Fig. \ref{fig:contextual} shows that points in a context-wise outlier represent a phenomenon that is anomalous with respect to a reference frame.
Identifying the anomaly accurately requires selecting the size of the reference frame carefully.
This study sheds light on the detection challenges introduced by context-wise outliers.

\subsection{Context-wise Outlier Sub-Categories}

\begin{figure}[t!]
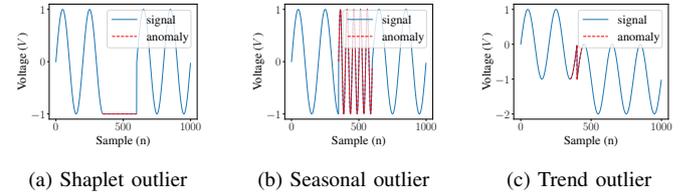

     \centering
     \begin{subfigure}[b]{0.3\columnwidth}
         \centering
         {\resizebox{\textwidth}{!}{\input{"outlier_taxonomy/shapelet_outlier.pgf"}}}
         \caption{Shaplet outlier}
         \label{fig:shaplet}
     \end{subfigure}
     \hfill
     \begin{subfigure}[b]{0.3\columnwidth}
         \centering
         {\resizebox{\textwidth}{!}{\input{"outlier_taxonomy/seasonal_outlier.pgf"}}}
         \caption{Seasonal outlier}
         \label{fig:seasonal}
     \end{subfigure}
     \hfill
     \begin{subfigure}[b]{0.3\columnwidth}
         \centering
         {\resizebox{\textwidth}{!}{\input{"outlier_taxonomy/trend_outlier.pgf"}}}
         \caption{Trend outlier}
         \label{fig:trend}
     \end{subfigure}
        \caption{Context-wise outlier taxonomy (adapted from \cite{lai2021revisiting}).}
        \label{fig:contextual-outliers}
        \vspace{-0.4cm}
\end{figure}

The context-wise outlier subset of anomalies represents various anomalous sub-sequences of the data in a given context. 
Since context-wise outliers are the focus of this study, it is possible to further classify this outlier type into three sub-categories, as illustrated in Fig. \ref{fig:contextual-outliers} and summarized as follows.

\subsubsection{Shaplet}
Shaplet outliers are classified by a shaplet or pattern that significantly differs from the normal data pattern.
This outlier type classifies abrupt faults in a system and constitutes an important outlier type for this study.

\subsubsection{Seasonal}
Seasonal outliers are classified by an increased or decrease pattern frequency during a specific time period.
Identifying seasonal outliers is important in understanding specific phenomena, e.g., a spike in web traffic related to a major holiday.
Another example is an increased demand in residential electrical demand because of a large televised sporting event.

\subsubsection{Trend}

Trend outliers are classified by a sub-sequence of the dataset that modifies the underlying distribution of the data.
Trend outliers are present in certain faults in the Power Electronic Converter (PEC) dataset under study.

\section{Detection Approaches} \label{three}

\subsection{Matrix Profile}

The Matrix Profile algorithm computes the distances between neighboring points in a dataset.
It exhibits a variety of advantages, including speed and generalizability to a variety of problem domains.
Authors in \cite{yeh2016matrix-profile-1} explain that in the algorithm, the results for each point or sub-sequence are stored in a vector and the combination of all the sub-sequences forms the overall matrix.
The conventional algorithm utilizes the Euclidean method to determine point-wise distance; other distance calculations can also be utilized.
Z-normalization technique is traditionally used to scale the data;
however, normalization can be modified or omitted depending on the specific dataset requirements.

Using a fixed ($m$) sized window, the algorithm computes the nearest-neighbor distances compared to the entire data stream. To select an appropriate window size ($m$), it is important to consider the granularity of the studied anomalies. A large windows size would detect and identify significant disturbances, like sensor failure, whereas a small window size would identify more localized disturbances, like a voltage sag, in a signal.

The results of the computation of the nearest-neighbor distances and the index of the closest neighbors are stored in order of closeness in a new index. In particular, the algorithm steps are outlined as follows \cite{matrix-profile-intro}:
\begin{enumerate}
    \item For each point in the window ($m$), compute the distance to the nearest neighbor against the entire data set.
    \item Exclude identical or nearly identical matches to prevent inaccuracy.
    \item Update the distance matrix with the new closest neighbor distance.
    \item Set the position matrix with the index position of the new closest neighbor.
\end{enumerate}

Using the Matrix Profile technique, it is easy to extract information, like motifs or repeated patterns, from the dataset.
More importantly, discords which represent anomalies can be discovered from a data stream.

\subsection{Deep Learning}
Recent emergence of transformer deep learning methods based on attention~\cite{vaswani2017attention} reveal promising results not only in the field of natural language processing~\cite{devlin2018bert}, but also in the processing of the tabular~\cite{arik2021tabnet} and time-series~\cite{lim2019temporal} data. Transformers were shown to be able to outperform the Gradient Boosting Decision Tree (GBDT) methods and ResNet-like architectures~\cite{NEURIPS2021_9d86d83f, klambauer2017self}. In this work, we focus on providing  complementary results to the Matrix Profile algorithm, by applying transformers for unsupervised anomaly detection~\cite{xu2022anomaly}. An anomaly-attention mechanism is used to compute the association discrepancy to distinguish between normal and abnormal (anomalous) data points. Adjusted implementation code of the authors\footnote{\url{https://github.com/5uperpalo/Anomaly-Transformer_FIREMAN}} and related datasets can be found on the FIREMAN project GitHub pages\footnote{\url{https://github.com/5uperpalo/FIREMAN-project/}}.

\section{Applications in Power Electronics} \label{four}

Transitioning from conventional power systems to power electronics-dominated grids (PEDG) has increased demand for grid-forming converters (GFM) to facilitate operational reliability.
GFMs have made significant progress in recent years to expedite stability under different grid conditions, but their operation during faults or large signal disturbances still remains a challenge.
Authors in \cite{trainsient-stability-9523750} note that GFMs handle a significantly smaller percentage of over-current (usually only 20\%) compared to synchronous generators (SGs) which can handle seven times their nominal current.
This makes fault detection for GFMs critical to maintain synchronization with the grid.
Since the network infrastructure of power systems keeps expanding, it is important to identify these faults accurately under varying grid parameter uncertainties.

\begin{figure}[t!]
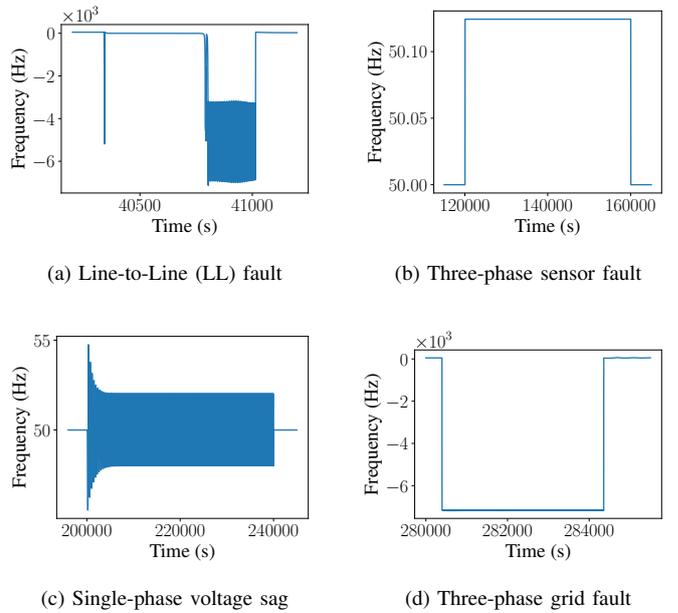

    \centering
    \begin{subfigure}[b]{0.475\columnwidth}
        \centering
        \resizebox{\textwidth}{!}{\input{"2_pec_sim/base_sig_f_c_f1.pgf"}}
        \caption{Line-to-Line (LL) fault}
        \label{fig:ll-fault}
    \end{subfigure}
    \hfill
    \begin{subfigure}[b]{0.475\columnwidth}
        \centering
        \resizebox{\textwidth}{!}{\input{"2_pec_sim/base_sig_f_c_f2.pgf"}}
        \caption{Three-phase sensor fault}
        \label{fig:three-phase-sensor-fault}
    \end{subfigure}
    \vskip\baselineskip
    \begin{subfigure}[b]{0.475\columnwidth}
        \centering
        \resizebox{\textwidth}{!}{\input{"2_pec_sim/base_sig_f_c_f3.pgf"}}
        \caption{Single-phase voltage sag}
        \label{fig:single-phase-voltage-sag}
    \end{subfigure}
    \hfill
    \begin{subfigure}[b]{0.475\columnwidth}
        \centering
        \resizebox{\textwidth}{!}{\input{"2_pec_sim/base_sig_f_c_f4.pgf"}}
        \caption{Three-phase grid fault}
        \label{fig:three-phase-grid-fault}
    \end{subfigure}
    \caption{PEC dataset fault visualization.}
    \vspace{-0.3cm}
    \label{fig:pec-faults-overall-fig}
\end{figure}

This study examines four faults in the PEC dataset: line-to-line (LL) fault, three-phase sensor fault, single-phase voltage sag and three-phase grid fault. It is worth noting that the location of these faults can be determined from Fig. \ref{fig:sys_model}.
The frequency [$f_c$] of the system throughout various fault conditions is used for detection.
Each fault has significantly different characteristics and magnitude.
The faults explained in this section are illustrated in Fig. \ref{fig:pec-faults-overall-fig}.








\subsection{Matrix Profile}
\begin{figure}[t!]
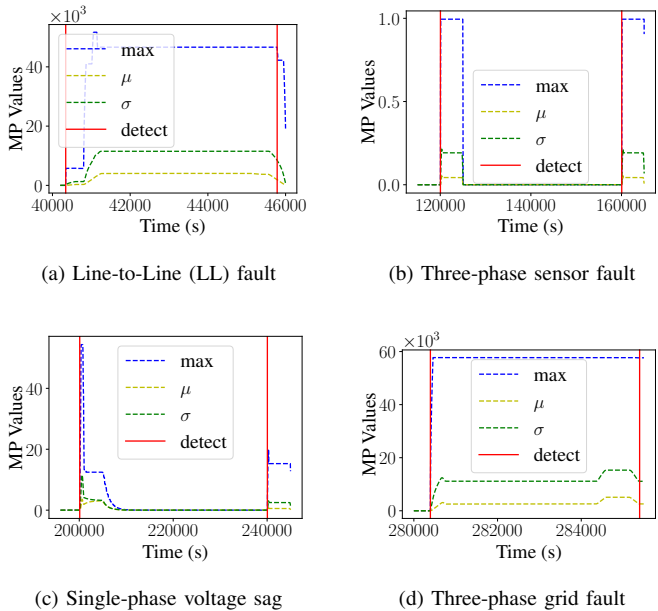

    \centering
    \begin{subfigure}[b]{0.475\columnwidth}
        \centering
        \resizebox{\textwidth}{!}{\input{"2_pec_sim/mp_hist_f_c_f1.pgf"}}
        \caption{Line-to-Line (LL) fault}  
        \label{fig:mp-ll-fault}
    \end{subfigure}
    \hfill
    \begin{subfigure}[b]{0.475\columnwidth}  
        \centering 
        \resizebox{\textwidth}{!}{\input{"2_pec_sim/mp_hist_f_c_f2.pgf"}}
        \caption{Three-phase sensor fault}  
        \label{fig:mp-three-phase-sensor-fault}
    \end{subfigure}
    \vskip\baselineskip
    \begin{subfigure}[b]{0.475\columnwidth}   
        \centering 
        \resizebox{\textwidth}{!}{\input{"2_pec_sim/mp_hist_f_c_f3.pgf"}}
        \caption{Single-phase voltage sag}    
        \label{fig:mp-single-phase-voltage-sag}
    \end{subfigure}
    \hfill
    \begin{subfigure}[b]{0.475\columnwidth}   
        \centering 
        \resizebox{\textwidth}{!}{\input{"2_pec_sim/mp_hist_f_c_f4.pgf"}}
        \caption{Three-phase grid fault}     
        \label{fig:mp-three-phase-grid-fault}
    \end{subfigure}
    \caption{PEC dataset fault detection using Matrix Profile algorithm.}  
    \label{fig:mp-pec-faults-overall-fig}
\end{figure}

\begin{figure}[t!]
    \centering
    \resizebox{\columnwidth}{!}{\input{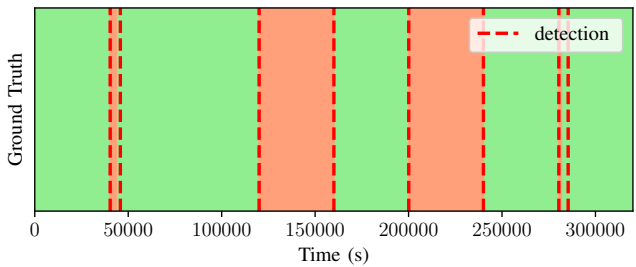}}
    \vspace{-0.7cm}
    \caption{PEC ground truth comparison [Normal: green, Anomaly: red].}
    \vspace{-0.3cm}
    \label{fig:pec_outliers}
\end{figure}

Fig. \ref{fig:mp-pec-faults-overall-fig} depicts a zoomed in window of the matrix profile values during each fault type examined in this experiment.
A description of the characteristics of these fault types can be found in Section \ref{two}.
The detector is shown to behave robustly and accurately for all fault types presented.
Fig. \ref{fig:pec_outliers} shows that the detector was able to accurately determine the start and end of each anomaly (i.e., 100\% detection rate) with no false positives (i.e., 0\% error rate).
It is noted that detecting the ending of an anomaly sometimes takes more time than detecting the start.
 
By nature, anomalies are rare, therefore interpreting the detection and false positive rates in a traditional way is not advisable.
In our case, the detector was able to successfully detect all four faults with zero false positives in the PEC dataset.
For the power grid use cases, it is important to detect anomalies quickly so that remedial action can be taken promptly.
Interestingly, the detector performed well when coping with anomalies of dramatically different characteristics and magnitude.
Each anomaly in the dataset is shaped differently, and the algorithm accurately detected the start and end of each one.
This demonstrates the robustness and scalable features of the algorithm, which was also designed to run in real-time and was simulated as such.
Each datapoint arrived in a simulated time series, and the algorithm was not aware of subsequent datapoints.
It utilizes a sliding window technique; therefore it fits in a fixed memory window.
Additionally, the computations are efficient and scale well with the window size; this offers strong potential in a real-time monitoring context.

\subsection{Anomaly-Transformer}
In Table~\ref{anomaly_transformer-tab}, we provide the detection performance summary per fault type using the anomaly-transformer in the PEC dataset. For model evaluation, we used full datasets per fault consisting of 13 features, including the frequency [$f_c$] used by the Matrix Profile algorithm. Training and testing parameter values are included in the provided GitHub repositories. From the results, we can observe that in the default configuration, the model was able to identify all anomalies. We also note that additional tuning is needed to decrease the number of false positives in the case of three-phase grid fault. The Matrix Profile is shown to outperform the model in our use case, but further testing with multi-feature anomalies and faults would be needed to confirm this observation. 

\begin{table}[h!]
\centering
\begin{tabular}[b]{|l|c|c|c|c|}
    \hline
	\textbf{Fault} & \textbf{Accuracy} & \textbf{Precision} & \textbf{Recall} & \textbf{F-score} \\ \hline
    LL fault & 0.990 & 0.860 & 1.000 & 0.925 \\ \hline
    Single-phase voltage sag & 0.998 & 0.998 & 1.000 & 0.999 \\ \hline
    Three-phase grid fault & 0.985 & 0.630 & 1.000 & 0.773 \\ \hline
    Three-phase sensor fault & 0.997 & 0.996 & 1.000 & 0.998 \\ \hline
\end{tabular}
\caption{Anomaly-transformer per fault performance report.}
\vspace{-0.3cm}
\label{anomaly_transformer-tab}
\end{table}

\section{Conclusions} \label{five}

Real-time detection and classification has significant implications in grid cyber-security and reliability.
Attacks against the power grid are becoming more sophisticated, and it is essential to discern whether the system is under attack or experiencing a fault condition.
Furthermore, it is important to know whether a fault is occurring so that automated or manual corrective actions can be performed, to protect system components and ensure maximum system reliability and uptime.
In this paper, we have presented different methods for anomaly detection while proposing an effective approach to identify faults in the operation of power electronic converters in PEDGs.

As future work, we plan to extend the proposed approach to a real-time dynamic detector.
There is currently a MATLAB power grid control and monitoring interface for a real system, and in order to integrate with it, a module for MATLAB must be created.
This would involve creating an implementation of the Matrix Profile algorithm in MATLAB, and utilizing it in tandem with the control system.
With this setup, it would be possible to test the real-time detection capabilities of the algorithm.
Once implemented, it would be possible to couple it with a classification algorithm to attempt to determine the type of fault.
If a fault was determined, the appropriate corrective actions could be performed depending on the classification.

\section*{Acknowledgment}
This work is partly supported by the FIREMAN project CHIST-ERA-17-BDSI-003 funded by the Spanish National Foundation (PCI2019-103780), the Academy of Finland (AoF; n.326270) and the Greek General Secretariat of Research and Technology; also by (1) AoF via  EnergyNet fellowship n.321265/n.328869/n.352654 and X-SDEN project n.349965, and (2) Baltic-Nordic Energy Research programme via Next-uGrid project n.117766. This work has been also funded by the "Ministerio de Asuntos Económicos y Transformación Digital" and the European Union-NextGenerationEU in the frameworks of the "Plan de Recuperación, Transformación y Resiliencia" and of the "Mecanismo de Recuperación y Resiliencia" under references TSI-063000-2021-39/40/41.
%

\bibliographystyle{ieeetr}
\bibliography{References}

\begin{thebibliography}{10}

\bibitem{he2020}
J.~He, Q.~Yang, and Z.~Wang, ``On-line fault diagnosis and fault-tolerant
  operation of modular multilevel converters — a comprehensive review,'' {\em
  CES Transactions on Electrical Machines and Systems}, vol.~4, no.~4,
  pp.~360--372, 2020.

\bibitem{bec2021}
F.~Becker, E.~Jamshidpour, P.~Poure, and S.~Saadate, ``Fault detection and
  localization for t-type converter,'' in {\em 2021 IEEE International
  Conference on Environment and Electrical Engineering and 2021 IEEE Industrial
  and Commercial Power Systems Europe}, pp.~1--4, 2021.

\bibitem{zhou2018}
D.~Zhou and Y.~Tang, ``An online open-circuit fault diagnosis and fault
  tolerant scheme for three-phase ac-dc converters with model predictive
  control,'' in {\em 2018 International Power Electronics Conference
  (IPEC-Niigata 2018 -ECCE Asia)}, pp.~434--438, 2018.

\bibitem{lezana2006}
P.~Lezana, J.~Rodrguez, R.~Aguilera, and C.~Silva, ``Fault detection on
  multicell converter based on output voltage frequency analysis,'' in {\em
  IECON 2006 - 32nd Annual Conference on IEEE Industrial Electronics},
  pp.~1691--1696, 2006.

\bibitem{xin2017}
B.~Xin, T.~Wang, and T.~Tang, ``A deep learning and softmax regression fault
  diagnosis method for multi-level converter,'' in {\em 2017 IEEE 11th
  International Symposium on Diagnostics for Electrical Machines, Power
  Electronics and Drives (SDEMPED)}, pp.~292--297, 2017.

\bibitem{sun2018}
Q.~Sun, Y.~Wang, and Y.~Jiang, ``A novel fault diagnostic approach for dc-dc
  converters based on csa-dbn,'' {\em IEEE Access}, vol.~6, pp.~6273--6285,
  2018.

\bibitem{dong2021}
H.~Dong, F.~Chen, Z.~Wang, L.~Jia, Y.~Qin, and J.~Man, ``An adaptive
  multisensor fault diagnosis method for high-speed train traction
  converters,'' {\em IEEE Trans. Power Electron.}, vol.~36, no.~6,
  pp.~6288--6302, 2021.

\bibitem{Anomaly1}
K.~Gupta, S.~Sahoo, R.~Mohanty, B.~K. Panigrahi, and F.~Blaabjerg,
  ``Decentralized anomaly characterization certificates in cyber-physical power
  electronics based power systems,'' in {\em 2021 IEEE 22nd Workshop on Control
  and Modelling of Power Electronics}, pp.~1--6, 2021.

\bibitem{Anomaly2}
K.~Bhatnagar, S.~Sahoo, F.~Iov, and F.~Blaabjerg, ``Physics guided data-driven
  characterization of anomalies in power electronic systems,'' in {\em 2021 6th
  IEEE Workshop on the Electronic Grid}, pp.~1--6, 2021.

\bibitem{Anomaly3}
V.~S.~B. Kurukuru, M.~A. Khan, and S.~Sahoo, ``Cyber security in power
  electronics using minimal data -- a physics-informed spline learning
  approach,'' {\em IEEE Transactions on Power Electronics}, pp.~1--6, 2022.

\bibitem{kim2020}
S.-H. Kim {\em et~al.}, ``Fault detection method using a convolution neural
  network for hybrid active neutral-point clamped inverters,'' {\em IEEE
  Access}, vol.~8, pp.~140632--140642, 2020.

\bibitem{gan2020}
J.~Gan {\em et~al.}, ``Intelligent fault diagnosis with deep architecture,'' in
  {\em 2020 International Conference on Cyber-Enabled Distributed Computing and
  Knowledge Discovery (CyberC)}, pp.~272--275, 2020.

\bibitem{ye2020}
S.~Ye, J.~Jiang, J.~Li, Y.~Liu, Z.~Zhou, and C.~Liu, ``Fault diagnosis and
  tolerance control of five-level nested npp converter using wavelet packet and
  lstm,'' {\em IEEE Transactions on Power Electronics}, vol.~35, no.~2,
  pp.~1907--1921, 2020.

\bibitem{xai}
E.~Tjoa and C.~Guan, ``A survey on explainable artificial intelligence (xai):
  Towards medical xai,'' 2019.

\bibitem{GutierrezRojas2022}
D.~Gutierrez-Rojas, I.~T. Christou, D.~Dantas, A.~Narayanan, P.~H.~J. Nardelli,
  and Y.~Yang, ``Performance evaluation of machine learning for fault selection
  in power transmission lines,'' {\em Knowledge and Information Systems},
  vol.~64, pp.~859--883, Feb. 2022.

\bibitem{nardelli2022cyber}
P.~H. Nardelli, {\em Cyber-physical Systems: Theory, Methodology, and
  Applications}.
\newblock John Wiley \& Sons, 2022.

\bibitem{pedro}
J.~Rocabert, A.~Luna, F.~Blaabjerg, and P.~Rodríguez, ``Control of power
  converters in ac microgrids,'' {\em IEEE Transactions on Power Electronics},
  vol.~27, no.~11, pp.~4734--4749, 2012.

\bibitem{explaining-adversarial-examples}
I.~J. Goodfellow, J.~Shlens, and C.~Szegedy, ``Explaining and harnessing
  adversarial examples,'' 2014.

\bibitem{black-box-explainability}
S.~Sahoo, H.~Wang, and F.~Blaabjerg, ``On the explainability of black box
  data-driven controllers for power electronic converters,'' in {\em 2021 IEEE
  Energy Conversion Congress and Exposition (ECCE)}, pp.~1366--1372, 2021.

\bibitem{bayesian-weight-uncertainty-neural-networks}
C.~Blundell, J.~Cornebise, K.~Kavukcuoglu, and D.~Wierstra, ``Weight
  uncertainty in neural networks,'' 2015.

\bibitem{gal2016dropout}
Y.~Gal and Z.~Ghahramani, ``Dropout as a bayesian approximation: Representing
  model uncertainty in deep learning,'' in {\em Proceedings of the 33rd
  International Conference on International Conference on Machine Learning -
  Volume 48}, ICML'16, p.~1050–1059, JMLR.org, 2016.

\bibitem{SHAP-og-paper}
S.~Lundberg and S.-I. Lee, ``A unified approach to interpreting model
  predictions,'' 2017.

\bibitem{apriori}
R.~Agrawal {\em et~al.}, ``Fast algorithms for mining association rules,'' in
  {\em Proc. 20th Int. Conf. very large data bases, VLDB}, vol.~1215,
  pp.~487--499, Citeseer, 1994.

\bibitem{fpgrowth}
J.~Han, J.~Pei, and Y.~Yin, ``Mining frequent patterns without candidate
  generation,'' {\em ACM sigmod record}, vol.~29, no.~2, pp.~1--12, 2000.

\bibitem{brs}
T.~Wang, C.~Rudin, F.~Doshi-Velez, Y.~Liu, E.~Klampfl, and P.~MacNeille, ``A
  bayesian framework for learning rule sets for interpretable classification,''
  {\em The Journal of Machine Learning Research}, vol.~18, no.~1,
  pp.~2357--2393, 2017.

\bibitem{qarma1}
I.~T. Christou, E.~Amolochitis, and Z.~Tan, ``A parallel/distributed
  algorithmic framework for mining all quantitative association rules,'' {\em
  CoRR}, vol.~abs/1804.06764, 2018.

\bibitem{qarma2}
I.~T. Christou, ``Avoiding the hay for the needle in the stack: Online rule
  pruning in rare events detection,'' in {\em 2019 16th International Symposium
  on Wireless Communication Systems (ISWCS)}, pp.~661--665, 2019.

\bibitem{lai2021revisiting}
K.-H. Lai, D.~Zha, J.~Xu, Y.~Zhao, G.~Wang, and X.~Hu, ``Revisiting time series
  outlier detection: Definitions and benchmarks,'' in {\em Thirty-fifth
  Conference on Neural Information Processing Systems Datasets and Benchmarks
  Track (Round 1)}, 2021.

\bibitem{9217234}
C.~Kalalas and J.~Alonso-Zarate, ``Sensor data reconstruction in industrial
  environments with cellular connectivity,'' in {\em 2020 IEEE 31st Annual
  International Symposium on Personal, Indoor and Mobile Radio Communications},
  pp.~1--6, 2020.

\bibitem{yeh2016matrix-profile-1}
C.-C.~M. Yeh {\em et~al.}, ``Matrix profile i: All pairs similarity joins for
  time series: A unifying view that includes motifs, discords and shapelets,''
  pp.~1317--1322, 12 2016.

\bibitem{matrix-profile-intro}
T.~Marrs, ``Introduction to matrix profiles,'' 2019.

\bibitem{vaswani2017attention}
A.~Vaswani {\em et~al.}, ``Attention is all you need,'' {\em Advances in neural
  information processing systems}, vol.~30, 2017.

\bibitem{devlin2018bert}
J.~Devlin, M.-W. Chang, K.~Lee, and K.~Toutanova, ``Bert: Pre-training of deep
  bidirectional transformers for language understanding,'' {\em arXiv preprint
  arXiv:1810.04805}, 2018.

\bibitem{arik2021tabnet}
S.~O. Ar{\i}k and T.~Pfister, ``Tabnet: Attentive interpretable tabular
  learning,'' in {\em AAAI}, vol.~35, pp.~6679--6687, 2021.

\bibitem{lim2019temporal}
B.~Lim, S.~O. Arik, N.~Loeff, and T.~Pfister, ``Temporal fusion transformers
  for interpretable multi-horizon time series forecasting,'' {\em arXiv
  preprint arXiv:1912.09363}, 2019.

\bibitem{NEURIPS2021_9d86d83f}
Y.~Gorishniy, I.~Rubachev, V.~Khrulkov, and A.~Babenko, ``Revisiting deep
  learning models for tabular data,'' in {\em Advances in Neural Information
  Processing Systems}, vol.~34, pp.~18932--18943, Curran Associates, Inc.,
  2021.

\bibitem{klambauer2017self}
G.~Klambauer, T.~Unterthiner, A.~Mayr, and S.~Hochreiter, ``Self-normalizing
  neural networks,'' {\em Advances in neural information processing systems},
  vol.~30, 2017.

\bibitem{xu2022anomaly}
J.~Xu, H.~Wu, J.~Wang, and M.~Long, ``Anomaly transformer: Time series anomaly
  detection with association discrepancy,'' in {\em International Conference on
  Learning Representations}, 2022.

\bibitem{trainsient-stability-9523750}
E.~Rokrok, T.~Qoria, A.~Bruyere, B.~Francois, and X.~Guillaud, ``Transient
  stability assessment and enhancement of grid-forming converters embedding
  current reference saturation as current limiting strategy,'' {\em IEEE Trans.
  Power Syst.}, vol.~37, no.~2, pp.~1519--1531, 2022.

\end{thebibliography}

\end{document}